\documentclass[twocolumn,preprintnumbers,superscriptaddress,endnote,nofootinbib,aps,prd]{revtex4}

\usepackage{datetime}
\usepackage{color,fancybox}

\usepackage{graphicx}
\usepackage{grffile}
\usepackage{slashed}
\usepackage{bbm}
\usepackage{footmisc}
\usepackage{multirow}
\usepackage{amsmath,amssymb}

\newcommand{\rig}{\rightarrow}

\newcommand{\be}{\begin{eqnarray*}}
\newcommand{\ee}{\end{eqnarray*}}
\newcommand{\eg}{{\emph{e.g.}}}
\newcommand{\ie}{{\emph{i.e.}}}
\newcommand{\gl}[1]{(\ref{#1})}

\newcommand{\bee}{\begin{eqnarray}}
\newcommand{\eee}{\end{eqnarray}}
\newcommand{\beeq}{\begin{equation}}
\newcommand{\eeeq}{\end{equation}}
\newcommand{\gev}{~{\rm{GeV}}}

\begin{document}

\preprint{FTUV-11-0620\;\;KA--TP--13--2011\;\;LPN11-32\;\;SFB/CPP--11--31}

\title{Precise predictions for W$\bf{\gamma\gamma}$+jet production at hadron colliders} 

\author{Francisco~Campanario}
\email{francam@particle.uni-karlsruhe.de}
\affiliation{Institute for Theoretical Physics, KIT, 76128 Karlsruhe, Germany}
\author{Christoph~Englert}
\email{c.englert@thphys.uni-heidelberg.de}
\affiliation{Institute for Theoretical Physics, Heidelberg University, 69120 Heidelberg, Germany}
\author{Michael~Rauch}
\email{rauch@particle.uni-karlsruhe.de}
\affiliation{Institute for Theoretical Physics, KIT, 76128 Karlsruhe, Germany}
\author{Dieter Zeppenfeld}
\email{dieter@particle.uni-karlsruhe.de}
\affiliation{Institute for Theoretical Physics, KIT, 76128 Karlsruhe, Germany}

\begin{abstract}
  In this letter we report on a calculation of $W^\pm
  \gamma\gamma$+jet production at next-to-leading order QCD. We
  include the leptonic decays of the $W$ and take into account all
  off-shell and finite width effects. This is the first computation
  which falls into the category of triboson+jet production at
  next-to-leading order QCD. In total we find sizable corrections with
  nontrivial phase space dependencies. Therefore, our results are
  important for phenomenological analyses such as the extraction of
  anomalous electroweak quartic couplings from inclusive hadron
  collider data.
\end{abstract} 


\maketitle        
%
%
%
\section{Introduction}
The production of multiple electroweak bosons is an important channel
to test experimental data, collected at both the Tevatron and the
Large Hadron Collider (LHC), against the theoretically
well-established Standard Model (SM) hypothesis. As the mechanism
of electroweak symmetry breaking is yet to be determined, precise
predictions of electroweak gauge boson production rates are necessary
to experimentally infer deviations from the electroweak coupling
pattern predicted by the SM.  It is well-known that computations of
production cross sections and differential distributions suffer from
severe theoretical shortcomings if they are limited to the
semi-classical (\ie~leading order, LO) approximation in perturbation
theory. The arising uncertainties intrinsic to fixed order
calculations are conventionally assessed by investigating variations
of renormalization and factorization scales, which are remnants of the
perturbative series' truncation.

The next-to-leading order (NLO) real emission contribution to the
hadronic cross section, however, can probe new partonic initial
states not present at LO, so that the LO scale uncertainty can
sometimes be totally misleading.  This is especially true for
processes which are characterized by a QCD singlet final state at LO,
\eg, electroweak triboson production
\cite{Hankele:2007sb,Bozzi:2010sj}.  For these channels the total NLO
correction factors are known to be particularly sizable,
$K=\sigma^{\rm{NLO}}/\sigma^{\rm{LO}}\sim 1.8$. The main reason for this
large correction is that the NLO corrections include new, large LO real
emission subprocesses initiated by gluons. Probing the protons' gluon
parton distribution at small momentum fractions with the real emission 
contribution, the large correction does not signal a breakdown of
perturbation theory, but strongly asks for perturbative improvements
of the one-jet-inclusive cross sections as a major contribution to the
full next-to-next-to-leading order cross sections. Similar
observations and conclusions hold for diboson \cite{Ohnemus:1991kk,Campbell:2011bn}
and diboson+jet production \cite{Dittmaier:2007th,
  wgamma,wgamma2, wz} (see also Ref.~\cite{Rubin:2010xp} for a related
discussion of $Z$+jets).

In this paper we report on the first calculation performed in the
context of triboson + jet production: 
$\stackrel{}{p}\stackrel{\text{\tiny(}-\text{\tiny)}}{p} \rig
\ell^-\bar\nu_\ell\gamma \gamma + {\rm{jet}}+X$ and $
\stackrel{}{p}\stackrel{\text{\tiny(}-\text{\tiny)}}{p}
\rig \ell^+\nu_\ell\gamma \gamma + {\rm{jet}}+X$. We include all
off-shell and finite width effects in our calculation,~\ie~we compute
the full matrix element at ${\cal{O}}(\alpha^4\alpha_s^2)$. For
convenience we will refer to these processes as
$W^\pm\gamma\gamma+{\rm{jet}}$ production.

We organize this paper as follows: Section~\ref{sec:elements} reviews
the technical details of the calculation and comments on the numerical
Monte Carlo implementation. In Sec.~\ref{sec:results} we discuss the
numerical results; we examine the cross sections' scale variations and
the impact of the QCD corrections on differential distributions.
Section~\ref{sec:conclusion} gives a summary of this work.

\section{Elements of the calculation}
\label{sec:elements}
The leading order ${\cal{O}}(\alpha^4\alpha_s)$ contribution to, \eg,
$\stackrel{}{p}\stackrel{\text{\tiny(}-\text{\tiny)}}{p}\rightarrow \ell^-\bar \nu_\ell\gamma \gamma + {\rm{jet}}$
is given by the subprocesses
\begin{subequations}
\label{eq:losubproc}
\bee
\label{eq:qqbar}
q\bar Q &\rightarrow & \ell^-\bar \nu_\ell\gamma \gamma + g \,, \\
\label{eq:gqbar}
g\bar Q &\rightarrow & \ell^-\bar \nu_\ell\gamma \gamma + \bar q  \,,\\
\label{eq:qg}
q g &\rightarrow & \ell^-\bar \nu_\ell\gamma \gamma + Q \,,
\eee
\end{subequations}
where $q=(d,s)$ and $Q=(u,c)$ denote the light down- and up-type quark
flavors, respectively.  Due to unitarity of the CKM matrix, any
dependence on the CKM matrix elements drops out for the flavor-summed
gluon-induced subprocesses of Eq.~\gl{eq:gqbar},~\gl{eq:qg}. These
subprocesses also dominate the hadronic cross sections at the LHC
because the protons are typically probed at small momentum fractions
for inclusive production. Consequently, a non-diagonal CKM matrix
decreases our LO result only at the per mill-level. This is well below
the residual (NLO) scale uncertainty and we therefore use a diagonal
CKM matrix in our calculation.  Furthermore, we do not include bottom
contributions to the hadronic $W\gamma\gamma$+jet production cross
section. They either involve top or bottom quarks in the final state
and are distinguishable experimentally by $b$ tagging.

\begin{figure}[!t]
\includegraphics[width=\columnwidth]{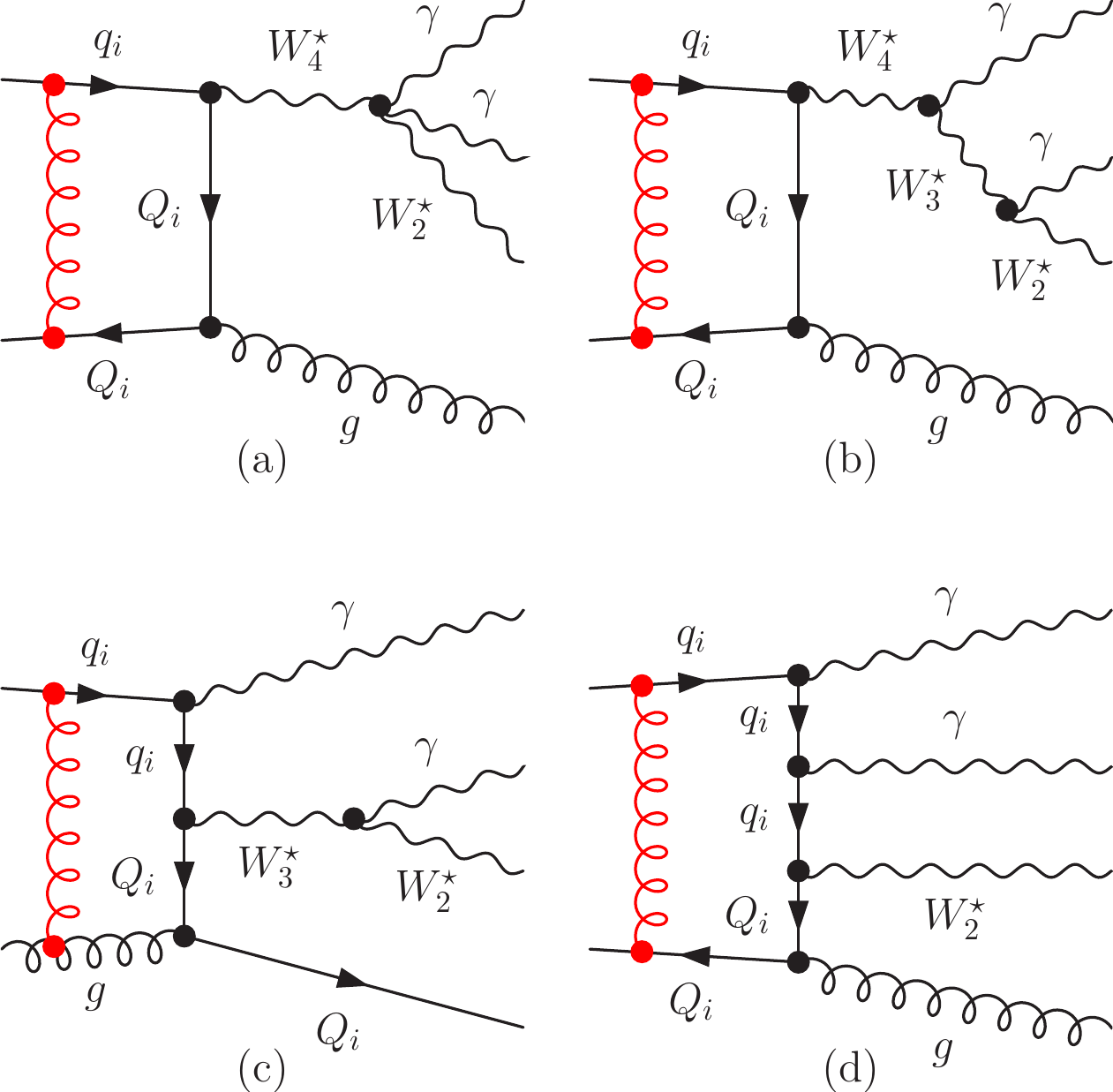}
\caption{\label{fig:waa} Selected topologies contributing to
  $W^-\gamma\gamma$+jet production at NLO; $i=1,2$ denotes the
  generation index. Note that we do not show the contributions
  $W^{\star}_n\rig e^-\bar\nu_e + (n-2)\gamma$, $2\leq n \leq 4$. These
  lead to identical configurations from the QCD point of view. The red
  loops indicate the virtual contributions, giving rise to topologies
  of up to boxes (a,b), up to pentagons (c) and up to hexagons (d).  }
\end{figure}

For the numerical implementation of the LO cross section we use
routines that are provided by the {\sc{Vbfnlo}} package
\cite{Arnold:2008rz} as the real emission contribution to $W\gamma
\gamma$ production at NLO QCD~\cite{Bozzi:2010sj}.  The hadronic part
of the amplitude is based on the spinor helicity formalism of
Ref.~\cite{Hagiwara:1988pp}, and the electroweak part of the amplitude
is provided by a cache system which employs {\sc{MadGraph}}-generated
{\sc{Helas}} routines~\cite{Alwall:2007st,Murayama:1992gi} using the
technique of ``\textsl{leptonic tensors}''~\cite{Jager:2006zc}. All
amplitudes of Eq.~\gl{eq:losubproc} are related by crossing symmetry,
and we show a sample of graphs contributing at NLO in
Fig.~\ref{fig:waa}.

\begin{figure}[!b]
\includegraphics[width=\columnwidth]{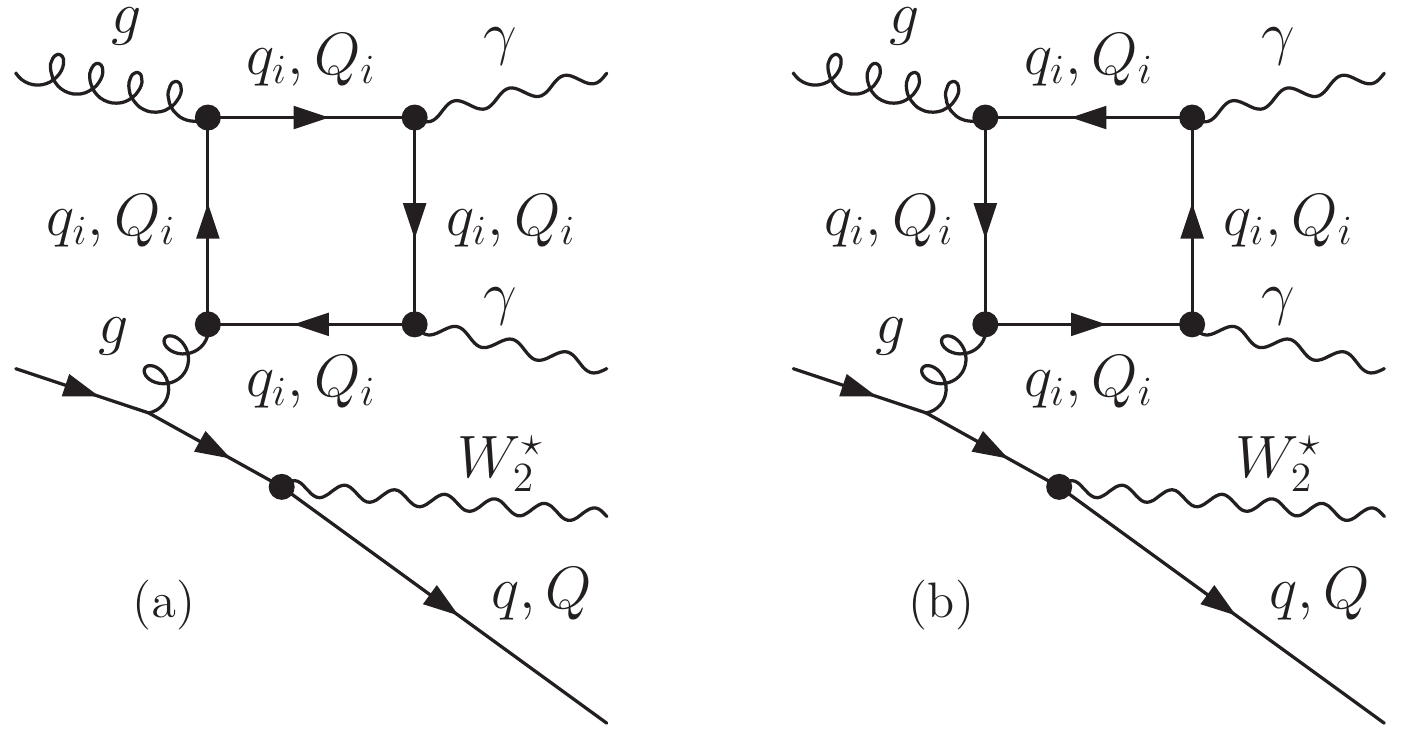}
\caption{\label{fig:waaferm} Fermion loop contributions to
  $W\gamma\gamma$+jet production at NLO; $i=1,2,3$ denotes the
  generation index. $q$ stands for all (anti)quark flavors of the
  incoming (anti)proton. Note that analogous $gg\gamma$ triangle
  contributions are forbidden by Furry's
  theorem. Not shown are topologies where the polarization vector
  $W^{\star}_2$ is attached to the initial state (anti)quark, and where
  the internal and external gluon are attached at opposite corners of
  the box.}
\end{figure}

For the virtual contributions we use the routines computed in
Ref.~\cite{Campanario:2011cs} which employ 
\textsc{FeynCalc} \cite{Mertig:1990an} and {\sc FeynArts}
\cite{Hahn:2000kx} in an in-house framework. We apply the effective current
approach described in, \eg,
Refs.~\cite{Bozzi:2010sj,wgamma,Figy:2003nv,Campanario:2011cs}, and 
combine all QCD corrections to sets of topologies with two, three, and
four attached (effective) gauge boson polarization vectors to standalone
numerical routines.\footnote{This approach allows to straightforwardly
  generalize our computation to models beyond the Standard Model as
  described in Refs.~\cite{wzano,wgamma2,Englert:2008wp}.}  The order
of the gauge bosons is thereby fixed and the full amplitude is
obtained by summing over all allowed permutations. The effective
$W^\star_n$ polarization vectors, see Fig.~\ref{fig:waa}, encode the
finite width effects in the fixed-width scheme of
Ref.~\cite{Denner:1999gp} and the off-shell contributions of the full
decays $W^\star_n \rig \ell\nu + (n-2)\gamma$, $2\leq n\leq4$, in a
straightforward way. To be more concrete, Figs.~\ref{fig:waa} (a) and
(b) contribute to a routine which includes all corrections up to boxes
(\ie~all corrections to $q\bar Q \rig W^\star_4 g$ with terms
proportional to $C_A$ and $C_F - \frac12 C_A$),
Fig.~\ref{fig:waa} (c) is part of a routine which includes corrections
up to pentagons ($\bar Qg \rig W^\star_3\gamma \bar q$), and
Fig.~\ref{fig:waa} (d) contributes to a routine  which also includes
hexagon diagrams ($q\bar Q \rig W^\star_2 \gamma \gamma g$). 
The fermion loop contributions are sketched in
Fig.~\ref{fig:waaferm} and were computed within the in-house
framework~\cite{Campanario:2011cs} and checked against an independent
implementation based on {\sc FeynArts}, {\sc FormCalc} and 
{\sc LoopTools}~\cite{Hahn:2000kx,Hahn:1998yk,Hahn:2006qw}. 

For the reduction of the tensor coefficients up to boxes we apply the
Passarino-Veltman approach of Ref.~\cite{Passarino:1978jh}, and for a
numerically stable implementation of 5- and 6-point coefficients we
use the Denner-Dittmaier scheme laid out in
Ref.~\cite{Denner:2002ii} with the setup and notation of 
Ref.~\cite{Campanario:2011cs}. This approach has turned out adequate in
a series of Feynman graph-based hexagon calculations, \eg,~in
Ref.~\cite{Ciccolini:2007ec}.  For completeness, we note that unitarity
cut-based methods have been demonstrated to be highly competitive in
Ref.~\cite{Bevilacqua:2009zn}.

\begin{figure*}[t!]
  \includegraphics[width=1\columnwidth]{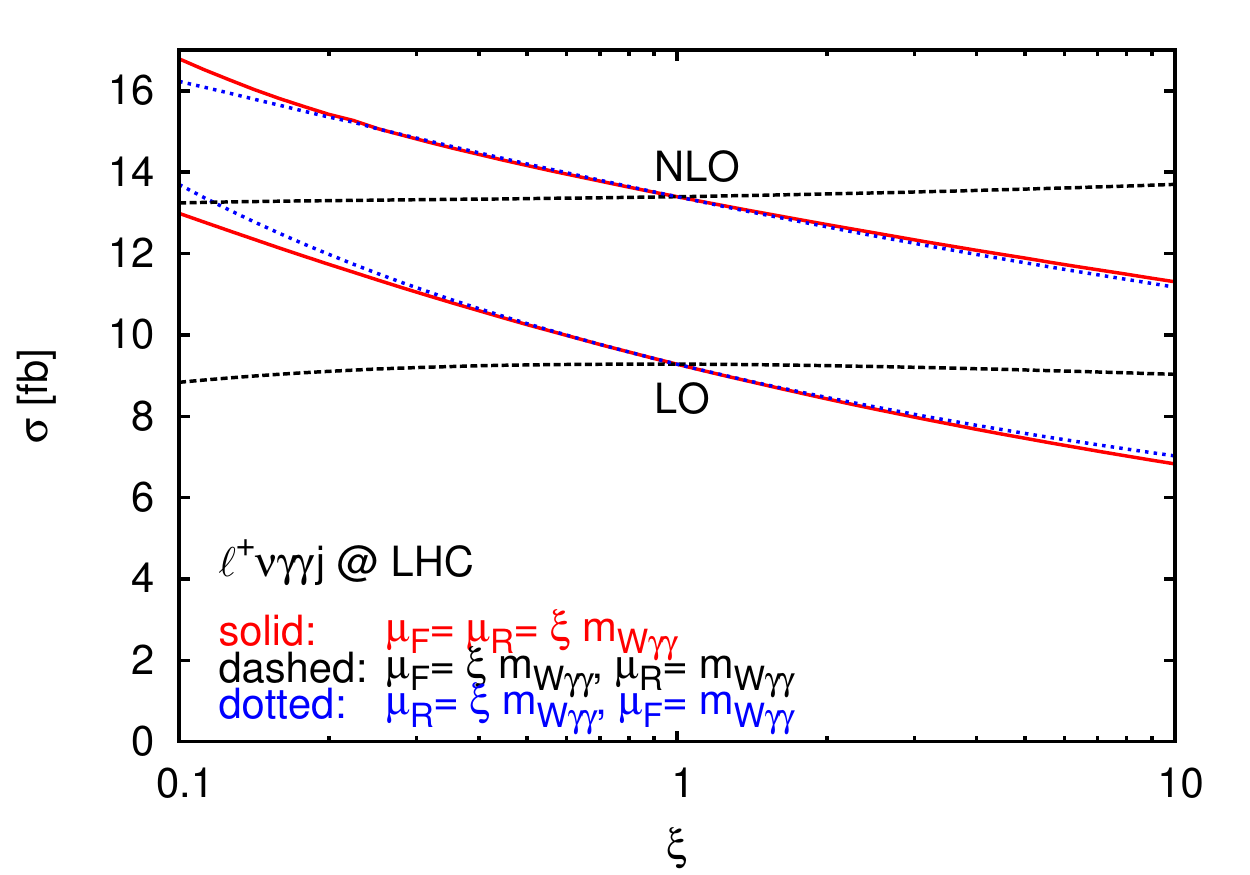}
  \hfill
  \includegraphics[width=1\columnwidth]{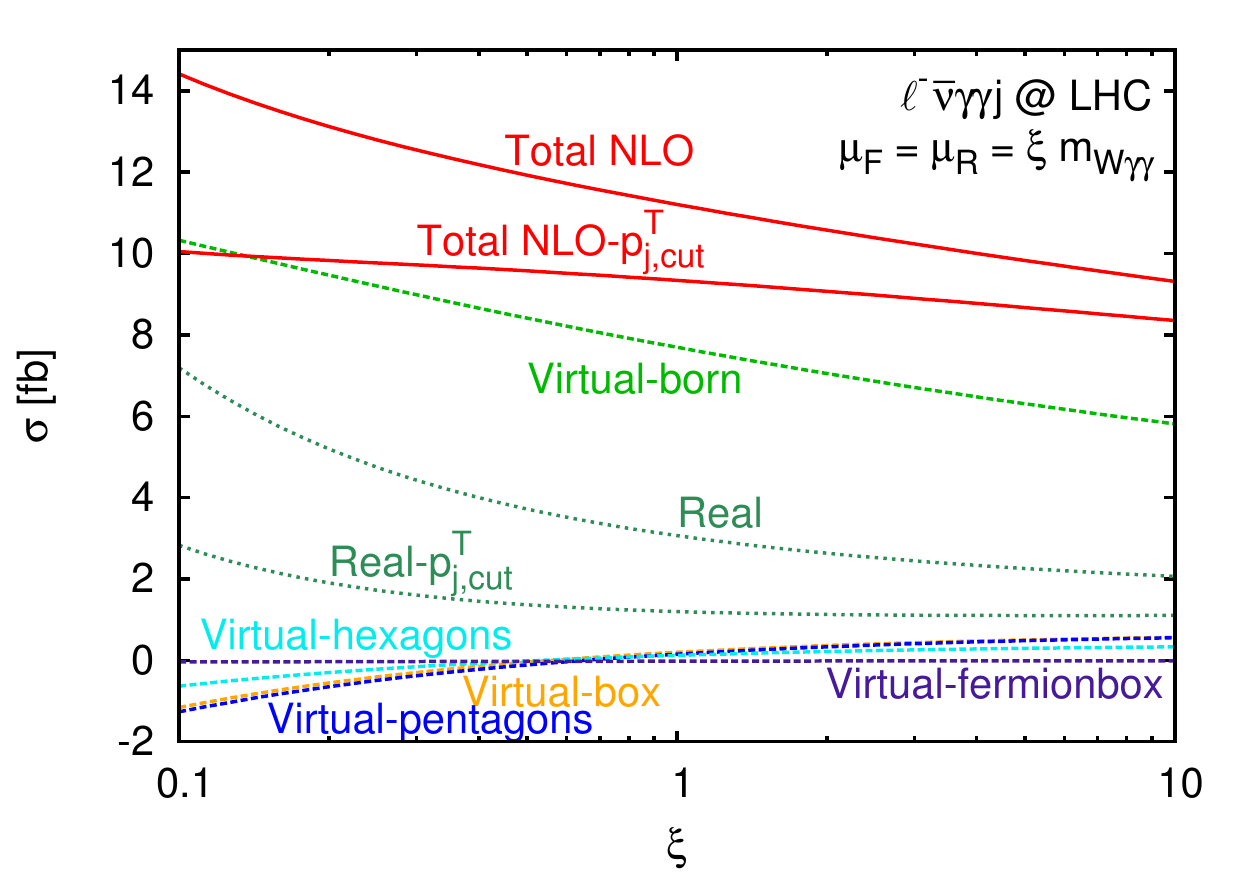}
  \caption{\label{fig:scalevar} Scale variation of the 
    $\ell^\pm \nu \gamma\gamma$+jet production cross sections
    at the LHC ($\ell$ = $e$, $\mu$). The cuts are
    described in the text and we take the invariant $W\gamma\gamma$ mass
    $m_{W\gamma\gamma}$ as central dynamical reference scale. 
    The left panel shows the variation of the LO and NLO 
    $W^+\gamma\gamma$+jet production cross sections when we change
    only the factorization scale, only the renormalization scale, or
    both jointly.
    For $W^-\gamma\gamma$+jet production the right panel shows the
    individual contributions to the NLO cross section, as discussed in
    the text. 
    Here we also present results where we have applied a veto
    on events with two identified jets having both a 
    transverse momentum larger than 50 GeV.
  }
\end{figure*}
%

The real emission contribution, $
\stackrel{}{p}\stackrel{\text{\tiny(}-\text{\tiny)}}{p} \rightarrow
\ell^-\bar \nu_\ell\gamma \gamma + 2\,{\rm{jets}}$, is based on the
existing implementations of Refs.~\cite{wgamma,wgamma2}.  We use the
Catani-Seymour dipole formalism \cite{Catani:1996vz} to numerically
regularize the soft and collinear QCD divergences and include finite
contributions after cancelling the infrared poles of the virtual
matrix element~\cite{Campanario:2011cs} to order
${\mathcal{O}}(\alpha^4\alpha_s^2)$.  The real
emission matrix element again implements the spinor-helicity formalism
of Ref.~\cite{Hagiwara:1988pp} and, analogous to the LO amplitude, the
implementation includes a cache system for the electroweak
$W^\star_n$ currents to minimize computation time. The implementation
of the dipoles is optimized to avoid redundancy, and the finite
collinear remainder, which is left after renormalizing the parton
densities, recycles the born-level matrix elements of the dipoles'
evaluation and is integrated over the real emission phase space
applying the phase space mappings of Ref.~\cite{Figy:2003nv}.

\begin{table}[!t]
\begin{tabular}{c c c c  c}
  \hline
  \hline
  & $\sigma^{\rm{LO}}$ [fb] & $\sigma^{\rm{NLO}}$ [fb] & $K=\sigma^{\rm{NLO}}/\sigma^{\rm{LO}}$ & \\ 
  \hline
  \hline
  $W^\pm\gamma\gamma$+jet & 1.191 &    1.754  & 1.47 & Tevatron \\
  \hline
  $W^+\gamma\gamma$+jet &  4.640  &   6.634  & 1.43 &  \multirow{2}{*}{LHC} \\
  $W^-\gamma\gamma$+jet &   3.803 &  5.644 &  1.48 & \\
\end{tabular}
\caption{\label{tab:xsecs} Total LO and NLO cross sections and
  $K$ factors for $\stackrel{}{p}\stackrel{\text{\tiny(}-\text{\tiny)}}{p}\rig
  e^-\bar\nu_e\gamma\gamma$+jet+$X$ and
  $\stackrel{}{p}\stackrel{\text{\tiny(}-\text{\tiny)}}{p}\rig
  e^+\nu_e\gamma\gamma$+jet$+X$ at the 
  Tevatron and at the LHC. The renormalization and factorization
  scales are chosen as $\mu_R=\mu_F=m_{W\gamma\gamma}$. Relative statistical
  and numerical stability errors are below the per mill level.}
\end{table}

\section{Numerical results}
\label{sec:results}

\begin{figure}[!t]
  \includegraphics[height=1.05\columnwidth]{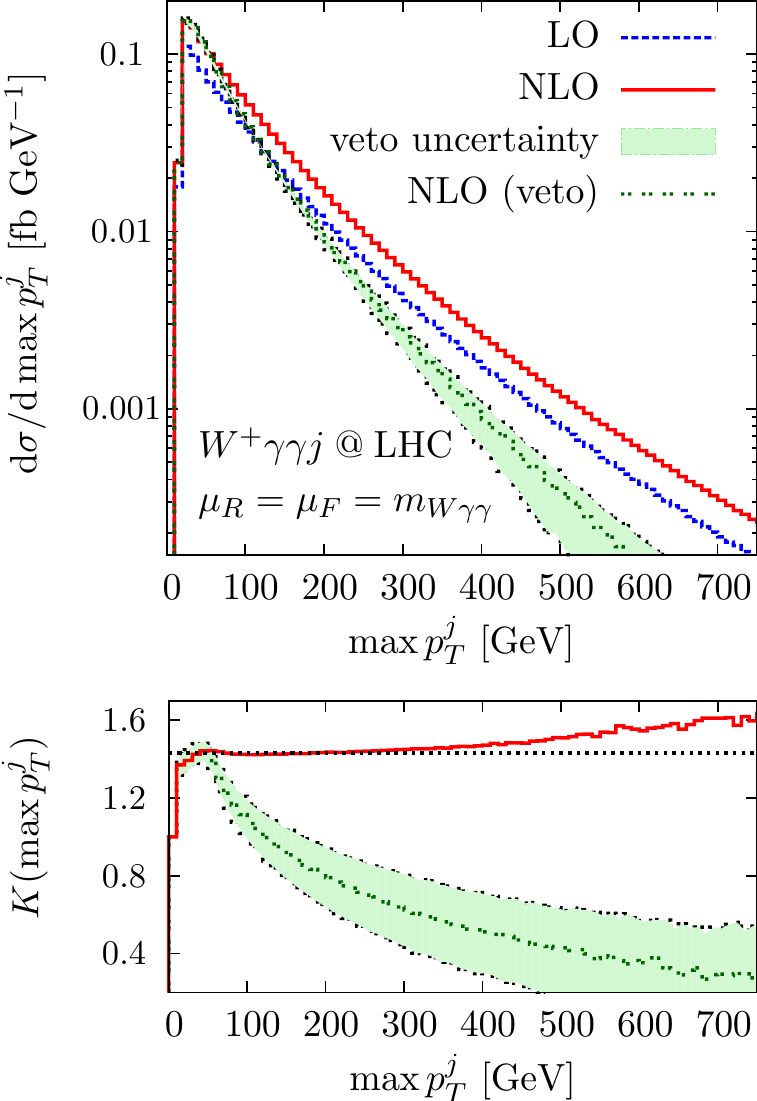}
  \caption{\label{fig:maxptj} Differential $\max p_T^j$ distribution
    for $W^-\gamma\gamma$+jet production. The chosen cuts and scales
    are described in text. The horizontal line in the lower panel
    displays the $K$ factor for total inclusive production, Tab.~\ref{tab:xsecs}.}
    \vspace{0.1cm}
\end{figure}

\begin{figure}[!b]
  \includegraphics[height=1.05\columnwidth]{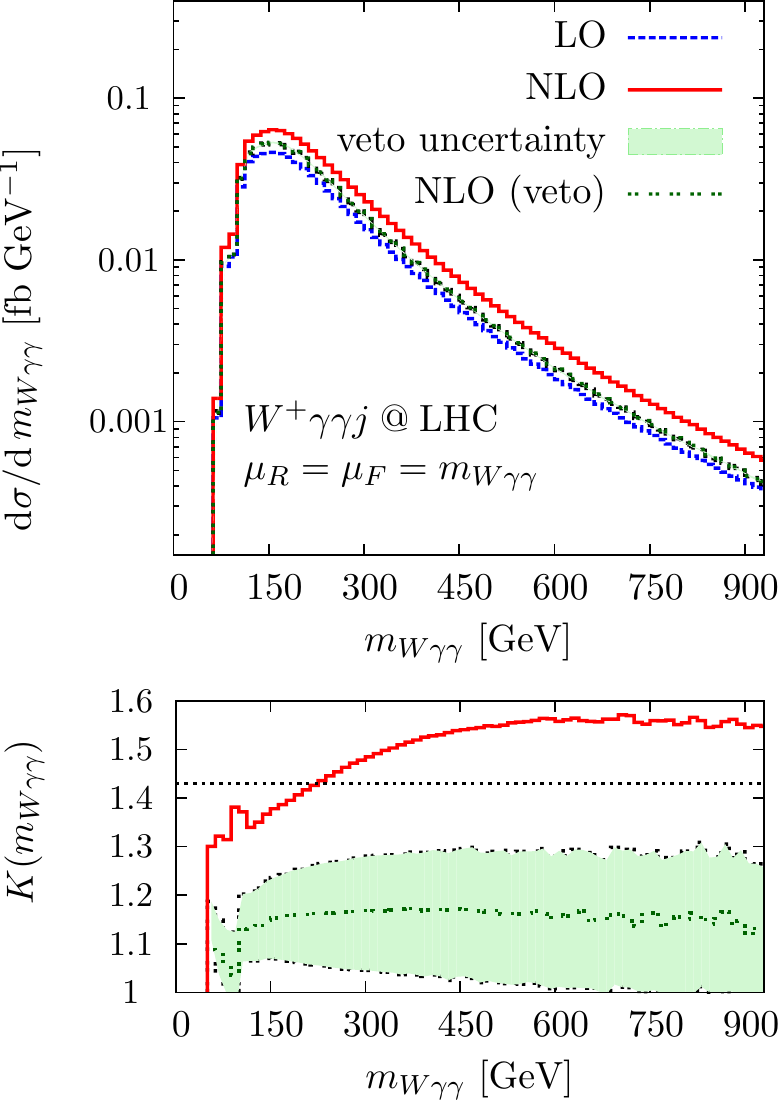}
  \caption{\label{fig:maxmaa} $W\gamma\gamma$ invariant mass
    distribution for $W^-\gamma\gamma$+jet production. The chosen cuts
    and scales are described in text. The horizontal line in the
    lower panel displays the $K$ factor for total
    inclusive production, Tab.~\ref{tab:xsecs}. }
\end{figure}

\begin{figure}[!t]
  \includegraphics[height=1.05\columnwidth]{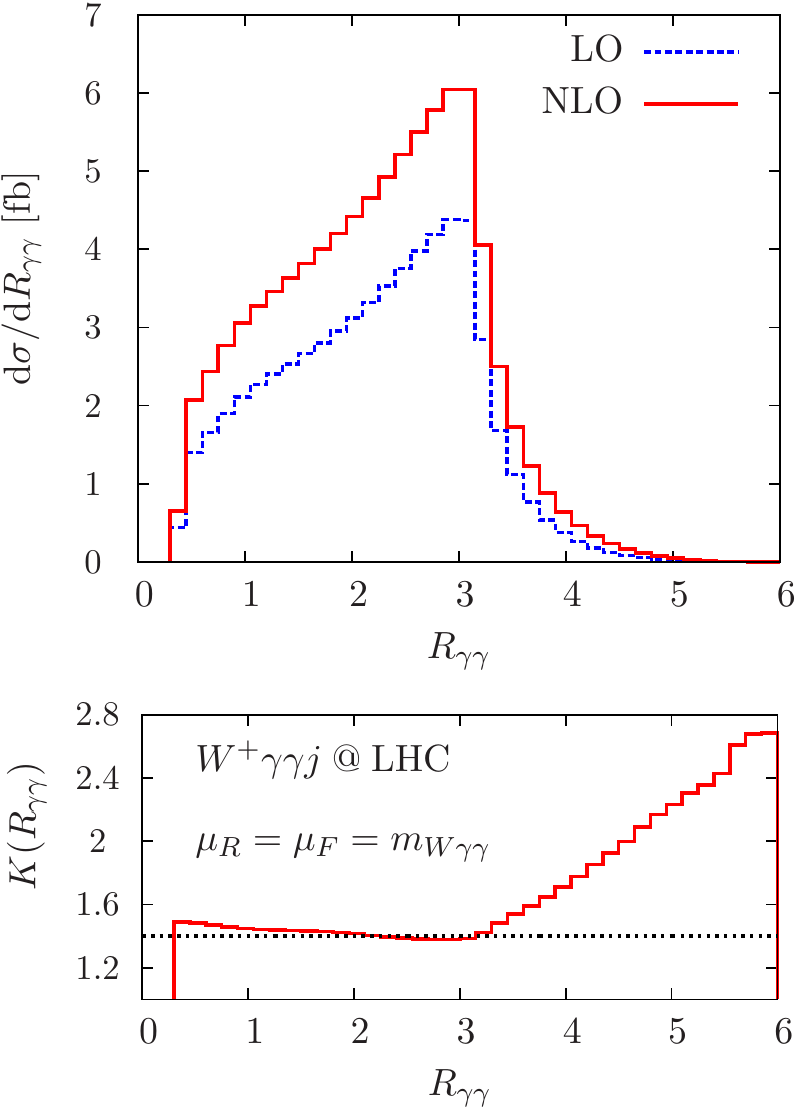}
  \caption{\label{fig:raa} Differential diphoton $R$ separation
    distribution for $W^-\gamma\gamma$+jet production. The chosen cuts
    and scales are described in text. The horizontal line in the lower
    panel displays the $K$ factor for total inclusive production,
    Tab.~\ref{tab:xsecs}. }
\end{figure}

We use CT10 parton distributions \cite{Lai:2010vv} with
$\alpha_s(m_Z)=0.118$ at NLO, and the CTEQ6L1
set~\cite{Pumplin:2002vw} with $\alpha_s(m_Z)=0.130$ 
at LO. We choose $m_{Z}=91.1876~\rm{GeV}$,
$m_{W}=80.398~\rm{GeV}$ and $G_F=1.16637\times
10^{-5}~\textnormal{GeV}^{-2}$ as electroweak input parameters and
derive the electromagnetic coupling $\alpha$ and the weak mixing angle
from Standard Model-tree level relations.  The center-of-mass energy
is fixed to $14~\rm{TeV}$ for LHC and $1.96~\rm{TeV}$ for Tevatron
collisions, respectively. We consider $W^\pm$ decays to the two light
lepton flavors,~\ie~for the distributions shown in
Figs.~\ref{fig:scalevar}-\ref{fig:raa} the decays $W\rightarrow
e\nu_e,\mu\nu_\mu$ have been summed, and we treat these
leptons as massless.

To study the impact of the QCD corrections on the process in detail,
we choose very inclusive cuts and a strictly isolated photon.  A naive
isolation criterion for the partons and the photon spoils infrared
safety by limiting the soft gluon emissions' phase space. Yet,
isolation is necessary to avoid non-perturbative jet-fragmentation
contributions, which would amount to the introduction of an additional
fragmentation scale to the problem. Instead, we apply the prescription
suggested in Ref.  \cite{Frixione:1998jh} (see also
Ref.~\cite{Baur:1993ir}), demanding \bee
\label{photonisolation}
\sum_{i, R_{i\gamma}<R} p_T^{{\rm{parton}}, i} \leq \frac{1- \cos
  R}{1- \cos \delta_0}\, p_T^\gamma \qquad \forall R\leq \delta_0,
\eee
\noindent where the index $i$ runs over all partons in a cone around
the photon of size $R$.  For the cut-off parameter, which determines
the QCD-IR-safe cone size around the photon, we choose $\delta_0=0.7$.
This is a rather large isolation compared to the experimental
resolution capabilities (\eg~Ref.~\cite{dc1}). Hence, the
phenomenological impact of the full jet-photon fragmentation is
expected to be small, in accordance with the results of
Refs.~\cite{wgamma2,Hoeche:2009xc,Baur:2010zf}.

\vfill

We cluster all final state partons with $|y_p|\leq 5$ to jets via the
$k_T$ algorithm \cite{Catani:1993hr} using a resolution parameter
$D=0.8$, adding the four momenta of clustered partons. The jets
are required to lie in the rapidity range $|y_j|\leq 4.5$ with
transverse momenta $p_T^{j}\geq 20~\rm{GeV}$. The photon and the
charged lepton are required to be hard and central, $p_{T}^\ell \geq
20 ~\rm{GeV}$ (10~GeV at the Tevatron), $p_{T}^\gamma \geq 20
~\rm{GeV}$ (10~GeV at the Tevatron), $|\eta_\ell|,|\eta_\gamma|\leq
2.5$, while being separated in the azimuthal angle-pseudorapidity
plane by $R_{\ell\gamma} = (\Delta \phi_{\ell\gamma}^2 +
\Delta\eta_{\ell\gamma}^2)^{1/2}\geq 0.4$.  For the separation of the
charged lepton from observable jets, we choose $R_{\ell j}\geq 0.4$
and we require $R_{\gamma\gamma}\geq 0.4$ for the diphoton separation.
Besides the photon-parton isolation criterion mentioned before, we also
require a separation between photons and identified jets of
$R_{\gamma j}\geq 0.7$.
The cross sections and total $K$ factors for a dynamical scale choice
$\mu_R=\mu_F=m_{W\gamma\gamma}$ are shown in Tab.~\ref{tab:xsecs}.
Here, $m_{W\gamma\gamma}$ denotes the $W^\pm\gamma\gamma$ invariant mass. 

The production cross section at the Tevatron is too small to be of
evident phenomenological importance when viewed in the light of
a total accumulated data set of $\sim 10 \text{ fb}^{-1}$ per
experiment.

We compute total $K$ factors of 1.43 (1.48) for
$W^+\gamma\gamma$+jet ($W^-\gamma\gamma$+jet) production at the
LHC. These values are quite typical for multiboson+jet production as
found in Refs.~\cite{Dittmaier:2007th, wgamma,wgamma2, wz}. 
The scale dependence of the $W^+\gamma\gamma j$ and $W^-\gamma\gamma j$
production cross sections turn out to be modest: when comparing 
$\mu_R=\mu_F=\xi m_{W\gamma\gamma}$ for $\xi=0.5$ and $\xi=2$, we find differences of
$10.8\%$ ($12.0\%$), respectively. 
The dependence is dominated by the renormalization scale, as can be seen
in Fig.~\ref{fig:scalevar}. This is a consequence of additional jet
radiation being important for the probed small gluon momentum fractions.
The right panel of
Fig.~\ref{fig:scalevar} also shows the effect of vetoing a second hard
jet. The real contributions become smaller and there is a cut-dependent
partial and accidental cancellation within the different contributions,
which should not to be taken as a stabilization of perturbation theory,
however (see also \cite{wgamma2,wz} and below).
For the numerical evaluation, we split the virtual contributions
into fermionic loops (Virtual-fermionbox, corresponding to the diagrams
sketched in Fig.~\ref{fig:waaferm}) and bosonic contributions with one,
two and three electroweak vector bosons attached to the quark line,
\ie{} Virtual-box, Figs.~\ref{fig:waa} (a) and (b), Virtual-pentagons,
Fig.~\ref{fig:waa} (c) or Virtual-hexagons, Fig.~\ref{fig:waa} (d),
respectively.
This procedure allows us to drastically reduce the time spent in
evaluating the part containing hexagon diagrams, which requires the
largest amount of CPU time. The bosonic contributions are not
individually QED gauge-invariant. However, this poses no problem since
for our choice of gauge (effectively, the Coulomb gauge in the lab-frame
is used for external photons) there are no sizable cancellations among
the different contributions.  
The fact that in Fig.~\ref{fig:scalevar} the different virtual
contributions share the same scale dependence corroborates our gauge
choice.

The phase space dependence of the QCD corrections is non-trivial and
sizable, as can be inferred from Figs.~\ref{fig:maxptj}-\ref{fig:raa},
where we again choose $\mu_R=\mu_F=m_{W\gamma\gamma}$.  Additional
parton emission redistributes the transverse momentum spectra. The
leading jet becomes slightly harder at NLO, an effect which is best seen
in the dynamical $K$ factor (ratio of NLO to  LO distribution) as shown
in the bottom panel of Fig.~\ref{fig:maxptj}. The effect on the 
$W\gamma\gamma$ invariant mass is even more pronounced.  While this
qualitative behavior is expected from kinematics, \eg~due to a photon
picking up the recoil from additional parton emission,
the quantitative result is very important. An excess in the
photon's transverse momentum or in the $W\gamma\gamma$ invariant mass 
at large values is easily misinterpreted as an effect of 
anomalous electroweak trilinear or quartic couplings 
\cite{wgamma2,  wz, Baur:1993ir} arising from new interactions beyond
the SM. 

The sizable impact of QCD corrections at large invariant masses is also
visible in the diphoton separation of Fig.~\ref{fig:raa}.  At large
values, when photons are highly separated in pseudorapidity, the
dynamical $K$-factor rises well above the average value of 1.43. 
This experimentally clean and
well-reconstructable configuration typically amounts to a large
momentum transfer in the quartic and trilinear vertices in
Figs.~\ref{fig:waa} (a) and (b) and therefore potentially accesses new
interactions at scales much larger than $m_W$.

In Figs.~\ref{fig:maxptj} and ~\ref{fig:maxmaa}, we also plot 
distributions with a veto on a second hard jet: no such jet with
$p_{T,j}>50$~GeV is allowed. It can be observed in
Fig.~\ref{fig:maxptj} that the vetoed contribution does not give a
sensible result for large values of $p_T$ of the harder jet, where large
logarithms involving $p_{T,\rm{cut}}$ as the other relevant scale
appear: changing values from $\mu_R=\mu_F=1/2 m_{W\gamma\gamma}$ to
$\mu_R=\mu_F=2 m_{W\gamma\gamma}$ increases the high $p_T$ tail of the
distribution by a factor 2.5 or more at NLO.
 In fact, in
Fig.~\ref{fig:maxptj} for $\max p_T^j\simeq 100\gev$, the scale-varied
distributions intersect, which is yet another clear indication of the
previously mentioned accidental stabilization of the vetoed cross
section, which is cut dependent.
In contrast, in Fig.~\ref{fig:maxmaa}, events with high invariant mass
can be generated, where the high mass has its origin purely in the
leptonic sector. These events can have two fairly soft jets which are not
cut away, which yields smaller variations in the $K$-factor, with
changes up to a factor 1.3 when increasing $\mu_R=\mu_F $ by a factor~4.

\bigskip
\section{Summary and Conclusions}
\label{sec:conclusion}
We have calculated the full NLO QCD corrections to the processes
$\stackrel{}{p}\stackrel{\text{\tiny(}-\text{\tiny)}}{p} \rig
\ell^-\bar\nu_\ell\gamma \gamma + {\rm{jet}}+X$ and
$\stackrel{}{p}\stackrel{\text{\tiny(}-\text{\tiny)}}{p}\rig \ell^+\nu_\ell\gamma
\gamma + {\rm{jet}}+X$.
All off-shell and finite width effects have been properly taken into
account. This is the first NLO computation which falls into the three
gauge boson-plus-jet category.

Quite typical for the multiboson+jet production modes we find large
total $K$ factors of order 1.4 for inclusive production, which are
driven by additional jet radiation being significant for our
inclusively chosen cuts. This enhancement is considerably larger than
naive expectations from a LO scale variation.  The corrections exhibit a
non-trivial phase space dependence, which could easily be
misinterpreted as non-Standard Model physics unless the differential
QCD corrections are properly included in experimental analyses.
\section*{Acknowledgments}
F.C. acknowledges partial support by FEDER and Spanish MICINN under
grant FPA2008-02878. CE thanks Michael Spannowsky for many helpful 
discussions. This research is partly funded by the Deutsche
Forschungsgemeinschaft under SFB TR-9 ``Computergest\"utzte
Theoretische Teilchenphysik'', and the Helmholtz alliance ``Physics at
the Terascale''.

\end{document}